\documentclass{aa}  
\usepackage{graphicx}
\usepackage{txfonts}
\begin{document}

\title{One blind and three targeted searches for (sub)millisecond pulsars}
   
\titlerunning{Searches for (sub)millisecond pulsars}

\author{E. Davoust \inst{1}
\and
G. Petit \inst{2}
\and
T. Fayard \inst{3}
}
\institute{
IRAP,
Universit\'e de Toulouse, CNRS, 
 14 Avenue Edouard Belin, F-31400 Toulouse, France\\
\email{emmanuel.davoust@ast.obs-mip.fr}
\and
BIPM, Pavillion de Breteuil, F-92312 S\`evres Cedex, France\\
\email{gpetit@bipm.org}
\and
 CNES/GRGS, 14 Avenue Edouard Belin, F-31400 Toulouse, France\\
\email{thierry.fayard@cnes.fr}
             }

\date{Received 2011; accepted }

% \abstract{}{}{}{}{} 
% 5 {} token are mandatory
\abstract
{Millisecond pulsars are very useful for determining the properties of neutron stars, for
testing General Relativity and for detecting gravitational waves. However, 
the number of known millisecond pulsars is very small compared to that of ordinary pulsars. 
}
{We conducted one blind and three targeted searches for millisecond and submillisecond pulsars
at radio frequencies. 
}
{The blind search was conducted within 3$^{\circ}$ of the Galactic plane
and at longitudes between 20$^{\circ}$ and 110$^{\circ}$. It takes 22073 pointings to cover this region,
and 5487 different positions in the sky (i.e. 25\% of the total) were actually observed.
The first targeted search was aimed at Galactic
globular clusters, the second one at 24 bright polarized and pointlike 
radiosources with steep spectra, and the third at 65 faint polarized and pointlike radiosources.
The observations were conducted at the large radiotelescope of Nan\c cay Observatory, 
at a frequency near 1400 MHz, the exact value depending on
the backend.  Two successive backends were used, first a VLBI S2 system, second
a digital acquisition board and a PC with large storage capacity sampling the signal at 50 Mb/s 
on one bit, over a 24-MHz band and in one polarization. 
The bandwidth of acquisition of the second backend was later increased to 48 MHz and the sampling rate to 100 Mb/s.
The survey used the three successive setups, with respective sensitivities of 3.5, 2.2, and 1.7 mJy.
The targeted-search data were obtained with the third setup and reduced with
a method based on the Hough transform, yielding a sensitivity of 0.9 mJy.
The processing of the data was done in slightly differed time by soft-correlation in all cases.
}
{No new short-period millisecond pulsars were discovered in the different searches.
To better understand the null result of the blind survey, we estimate the probability of detecting 
one or more short-period pulsars among a given Galactic population of synthetic pulsars with our setup: 
25\% for the actual incomplete survey and
79\% if we had completed the whole survey with a uniform nominal sensitivity of 1.7 mJy.
The alternative of surveying a smaller, presumably more densely populated, region with a higher
sensitivity would have a low return and would be impractical at a transit instrument.
}
{The null result of our blind survey is compatible with our present understanding of the
Galactic population of millisecond pulsars. In particular, there does not seem to be a
large population of fast-rotating millisecond pulsars.\\}
    

\maketitle
%
%________________________________________________________________

\section{Introduction}
\label{intro}

In the past fifteen years the timing of millisecond pulsars in the radio domain has
become an important field of research for pulsars and their emission
mechanisms. Important results have been obtained in neutron-star physics and
General Relativity (see Lattimer \& Prakash \cite{lattimer} and Lorimer \cite{lorimer08} for reviews).
It has also been shown that, for certain
pulsars, the stability of the rotation period is close to that of atomic
time scales for an averaging time of several years (e.g. Janssen et al. \cite{janssen}). However, this 
property generally does not hold when the observations extend over decades.
Nevertheless, assuming a good stability of the rotation, a
``pulsar" time scale may be calculated from a pulsar timing array
(Taylor \cite{taylor}, Petit \& Tavella \cite{petit}, Guinot \& Arias \cite{guinot}, Manchester \cite{M11}).

It is thus essential to discover new millisecond pulsars in order to improve on the
present results.  In addition to better statistics on a larger set of data,
one can hope to discover pulsars with a better stability of rotation.
Presently, it seems that old millisecond pulsars with short period (less than 15 ms)
and thus small period derivative are the most plausible candidates for a stable period.
Furthermore, the timing of a widely distributed sample of millisecond pulsars has the
potential to detect gravitational waves at nanohertz frequencies 
(van Haasteren et al. \cite{vanhaasteren}, Verbiest et al. \cite{verbiest}, Manchester \cite{M11}).
Finally, the discovery of submillisecond pulsars would revolutionize
fundamental physics, because the stiffest equations of state would then have
to be abandoned (Haensel et al. \cite{haensel}).

It is likely that only a small fraction of the Galactic population of millisecond
pulsars has so far been discovered.  One reason for a low detection rate
is that the sensitivity of many surveys decreases strongly with decreasing rotation period, 
mainly for high values of interstellar dispersion, because of insufficient sampling resolution in time and frequency. 
In particular, the searches at the turn of the century for submillisecond pulsars were unsuccessful
(D'Amico \cite{damico}, Crawford et al. \cite{crawford}, Edwards et al. \cite{edwards}).

This paper reports on a blind survey in the Galactic plane and three successive targeted 
searches conducted at the large radiotelescope of Nan\c cay Observatory for short-period 
millisecond pulsars (ranging from 0.65 to 30 ms).  The project was initiated in the 1990s
to search for short-period pulsars with a technique of soft-correlation of radio signals
developed for other applications.  At the time, the recording of radio signals at a very
high sampling rate and their subsequent treatment by soft-correlation techniques was
quite a challenge in view of the performances of standard computers, and was only 
being done in VLBI with dedicated instruments. The first years were devoted to writing 
software to read the VHS (and later SVHS) video tapes on which the data were initially stored and to
developing dedicated parallel computers based on transputers, and later on shark processors, 
to perform the actual soft-correlations, along with the necessary software.
The blind survey started in 1997, a time when
past surveys had searched only a fraction of the parameter space, because of the limited
sampling resolution in time and frequency. It was designed to be complementary to past and ongoing
surveys by only searching with high sampling resolution. When the blind survey proved less
successful than expected, targeted searches were meant to provide additional chances 
of finding pulsars with limited additional observing time.

The acquisition method is explained in
Sect.~\ref{acquisition}, the data processing in Sect.~\ref{processing}, and its validation in Sect.~\ref{validation}.
The blind survey and the three targeted searches are presented in Sect.~\ref{searches}. 
We put the null result of the blind survey in perspective by estimating in Sect.~\ref{mspopulation}
the chances of detecting short-period pulsars with our setup,
by way of simulations on a model for the Galactic population of millisecond pulsars.

%__________________________________________________________________
\section{Method of acquisition}
\label{acquisition}

The observations were obtained at the large 
radiotelescope\footnote{http://www.obs-nancay.fr/en/index.php/instruments/radio-telescope}  
of Nan\c cay Observatory.
This is a transit instrument with a beam of 3.6x22 arcmin
at zero declination and 1400 MHz.

In the initial stages of the project (1997-98), the data acquisition was
done with the VLBI S2 system and SVHS video tapes.
The signal from the receiver in two bands (1400 and 1410 MHz) and in both polarizations
was registered, totalling four channels, each 6.4 MHz wide. The sensitivity of
the system was determined from the mean flux density of a detectable pulsar
(7 sigmas). We measured a sensitivity of 3.5 mJy for 240 s of integration,
from the signal-to-noise ratio (SNR) of \object{PSR B1937+21} (Rougeaux et al. \cite{BR2000}).
We obtained 186 positions in the sky (hereafter points) of the blind survey
with this setup. Shortly thereafter, the radiotelescope closed for 
renovation of the receiver, and the French space agency discontinued
its support for this project, thus bringing observations and technical 
developments to a temporary halt.

In 2004 we replaced this system with a custom-made acquisition board that digitized
the signal on one bit before sending it to an electronic board ADLink PCI-7300A revB, 
which stored the data on the 120Gb hard disk of a PC running under Windows 2000 OS.
The system sampled
the signal at a rate of 50 Mb/s on one bit and a 24-MHz band, one polarization, one channel,
at the frequency 1370 MHz. One
channel of this system (24 MHz) was equivalent to the old one (4x6.4 MHz), but the new system 
was much simpler to operate.
We also confirmed the gain in sensitivity by a factor 2.2 of the new receiver
of the radiotelescope after 2000. Procedures allowed the automatic acquisition of the
data and their analysis in slightly differed time on PCs.
The computation of the autocorrelation function with small time samples (typically 32 $\mu s$) 
was done on that computer
before further treatment on more powerful computers (part of a cluster of 98 PC boards, and later two
fast dual-core computers operating under Linux OS). 
The integration time of a single pointing was set to 200 s and the corresponding sensitivity was 2.2 mJy.
We obtained 2829 points of the blind survey with this setup.

In July 2006, a new custom-made acquisition board was installed at the radiotelescope. The signal
was thereafter obtained at 1392 MHz, with a 48-MHz bandwidth and sampled at a rate of 100Mb/s,
which raised the sensitivity by a factor 1.4. The integration time of each pointing was reduced
to 160 s, so the sensitivity was 1.7 mJy.
All three targeted searches and 2542 points of the blind survey were done with this setup.

The functioning of the backend was regularly checked by observations of
\object{PSR B1937+21}, about twice a month.

%__________________________________________________________________
\section{Data processing}
\label{processing}

Our data processing scheme for 1-bit data has been described earlier (Rougeaux \cite{BR}, Rougeaux et al. \cite{BR2000}).
Here we recall the algorithm used to detect pulsar signal in a single observation (Sect.~\ref{proc_one_obs}).
This processing method is hereafter called ``survey mode".
We then present the algorithm used to increase the sensitivity by using multiple consecutive observations 
(Sect. \ref{proc_n_obs}), which we call ``cluster mode". The blind survey was reduced with the
first method, while the reobservations of pulsar candidates in the blind survey 
and the three targeted searches were reduced with the other one.

\subsection{Processing in ``survey mode"}
\label{proc_one_obs}

The first step of the processing is coherent folding of the signal.
Then a search in a 3-dimensional parameter space (period $P$, dispersion slope $D$, 
and acceleration parameter $A$) is performed.
The range of trial periods was initially 0.65 -- 3.0 ms. 
The lower limit corresponds to the fastest theoretically possible rotation period
for a uniformly rotating neutron star (Haensel et al. \cite{haensel}).  Following the recommendation
of the radiotelescope time allocation committee, we extended this range
to 0.65 -- 30 ms when we started the targeted searches at the end of 2006, at a low computational cost (about 25\%). 
We then also applied the extended range to the blind survey, thanks to the increased computational power available.

For a given set of trial values $P, D, A$, the folding process yields an SNR value $S(P,D,A)$. 
The typical numbers of trials  
are $N_P \simeq 10^6-2\times10^6$ for period, $N_D \simeq$ 100-200 for dispersion slope, 
$N_A <$ 10 for acceleration. 
In addition, for each trial, the signal is searched for in about ten positions in phase so 
that the search space has about $10^{10}$ elements. 
All SNR values higher than a given threshold 
(e.g. for an SNR threshold of 4.0, typically $10^{6}$ elements out of $10^{10}$) are kept for further analysis. 
Because the acceleration parameters are relatively few, 
the results are examined independently for each value of this parameter. 
The retained values $S_A(P,D)$ are then grouped by peaks 
(clusters of adjacent points in the plane $P-D$). 

Once this basic treatment is done, all signals with
maximum SNR higher than some threshold (typically 6.5) are extracted and examined.
Except when radio-interferences are present, the processing typically yields one 
such candidate peak every three to five observations. 
For the candidate peaks, the raw data are reanalyzed with slightly different trial values, 
centered on the candidate values, and with finer steps in period, dispersion slope and acceleration. 
Candidates that maintain the SNR for varying trial values (about one per 50-100 observations) are reobserved.

\subsection{Processing in ``cluster mode"}
\label{proc_n_obs}

For $N_O$ consecutive observations, the results are then $N_O$ series $S_i(P,D,A)$, with $i$ from 1 to $N_O$. 
For a pulsar in a binary system, 
$P$ may vary within each observation (i.e. the acceleration parameter is not zero), 
and $P$ will also vary between each observation, therefore a new parameter $P_d = P(i+1) - P(i)$ is needed. 
Obviously the parameters $A_i$ and $P_d$ are correlated. 
However, the $A_i$ are poorly determined in each observation, 
thus they cannot directly be used to connect candidate peaks between the $N_O$ consecutive observations, 
especially as the candidate peaks are quite numerous (all SNR greater than typically 4-4.5). 
On the other hand, the dispersion measure of a pulsar is the same in all observations. 
The processing is then broken into $N_D$ independent procedures, each concerning one 
trial dispersion slope, for which the matching of ($P, A, P_d$) is performed as follows.

The Hough transformation (Aulbert \cite{aulbert}) is used to transform the search in the plane ($T,P$), 
where $T$ represents time (i.e. the sequence of observations), into one in the plane ($P_d,P$). 
Assuming that the total time span is short enough for a pulsar period to vary linearly with time, 
in the first parameter space, a pulsar is represented by a line $P = P0_d.i + P0$ where $P0$ is 
the period at the reference epoch, and $P0_d$ the period derivative. 
The line representing a pulsar is to be found as one that fits the measurement results ($i, P_i$). 
In the second parameter space ($P_d,P$), the measurement results are transformed to lines $P = P_i - i.P_d$ 
and the pulsar is represented by a single point ($P0_d, P0$). 
The point representing the pulsar is found at an intersection of lines. 

With respect to this basic procedure (see a more detailed description in Aulbert, \cite{aulbert}), 
two additional features are introduced, as each measurement ($i, P_i$) comes with an SNR $S_i$ 
and an acceleration parameter $A_i$. First, the acceleration parameter is used to constrain the P$_d$ value; 
i.e., each measurement produces not a full line, but a segment in the ($P, P_d$) plane. 
Second, each segment is assigned $S_i$ as a weight so that each intersection of lines 
comes with the combined weight of the participating segments, 
in effect providing the detection SNR if the intersection of lines corresponds to a real pulsar.

The results $S(P0, P0_d$) for all dispersion slopes $D$ are then assembled and 
grouped by peaks (clusters of adjacent points in the plane $P-D$) 
and peaks with large SNR are examined.  The couple ($P0, P0_d$) 
for the maximum SNR provides the period and period derivative of the observed, or candidate, pulsar. 
An example of such a processing is given in Sect.~\ref{perfhough}. 

\begin{table*}
 \centering
  \caption{Known pulsars observed for checking the sensitivity
}
\label{tabletest}
\begin{tabular}{l|rrrrrr}
\hline
Pulsar&$P$&$DM$&flux&N&obsSNR&obs flux\\
&ms&pc cm$^{-3}$&mJy&&&mJy\\
\hline
\object{PSR B1620-26}&11.076&63&1.6&4&19&1.9 - 2.1\\
\object{PSR B1744-24}A&1.396&235&2.5&3&30-35&3.0 - 3.5\\
\object{PSR J1807-2459A}&3.059&134&1.1&3&8.4&0.8 - 0.9\\
\object{PSR B1821-24}&3.054&120&1.1&4&8&0.8 - 0.9\\
\object{PSR J2051-0827}&4.509&21&2.8&4&9&0.9 - 1.0\\
\hline
\end{tabular}
\end{table*}

\section{Theoretical performance}
\label{validation}

The point source efficiency at zero declination is about 1.6 K/Jy at 1400 MHz, and the system temperature is about 35 K, 
which leads to a 7-sigma theoretical detection threshold of 1.2 mJy for one 160-s observation at 100 Mb/s. 
The actual sensitivity is lower for several reasons: loss due to the 1-bit digitization, 
decrease in sensitivity with declination differing from 0, losses due to rounding in the folding process, etc. 

The sensitivity of the targeted search was tested by the detection of known faint pulsars
in globular clusters, in the framework of the targeted search in globular clusters (see Sect.~\ref{gc}).
Table~\ref{tabletest} gives the results for five pulsars, each observed with three or four successive 
integrations of 160 s (see more details in Sect. \ref{perfhough}).

\begin{figure}
\centering
\includegraphics[width=8.5 true cm]{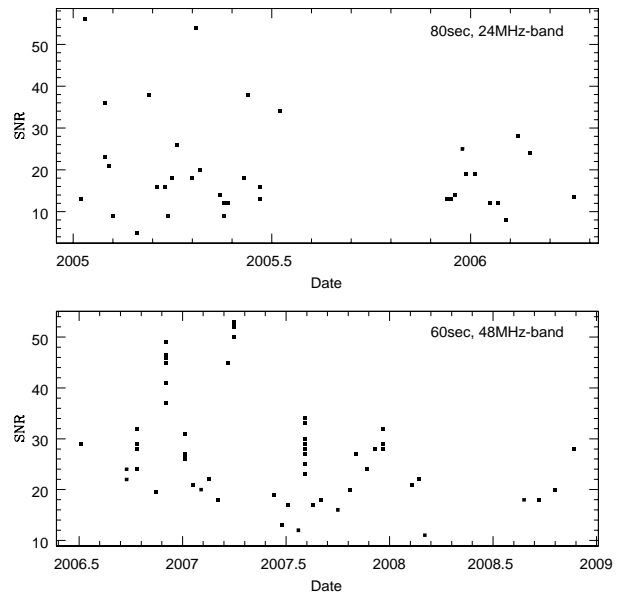}
\caption{Signal-to-noise ratio for single observations of \object{PSR B1937+21} and reduction in ``survey mode".
Top: 80-s pointings with second setup (24-MHz bandwidth).
There were no observations in the second semester of 2005 because of
a backend malfunction.
Bottom: 60-s pointings with third setup (48-MHz bandwidth).
These regular observations of the pulsar were made to check the
functioning of the backend and monitor the sensitivity of the setup.
}
\label{figsnr1937}
\end{figure}

\subsection {Performance validation for ``survey mode"}

We regularly observed the millisecond pulsar \object{PSR B1937+21} ($P$=1.557 ms, $DM=71 $cm$^{-3}$pc)\footnote{Most
data on known pulsars were taken from the ATNF pulsar catalog (http://www.atnf.csiro.au/research/pulsar/psrcat/);
see also Manchester et al. (\cite{M05})} 
in order to check the operation of the system and to obtain estimates of the achieved detection sensitivity. 
With the last setup, we obtained 99 60-s observations on 32 different days (see Fig.~\ref{figsnr1937}). 
The average detection SNR is 28.3 and the median value is 27.

Our results imply that \object{PSR B1937+21} would be detected with an SNR of  
46 (or 44 when using the median value) with an integration of 160 s. 
Assuming that the main pulse of \object{PSR B1937+21} is a source with a mean flux density of 10 mJy, 
we infer a 7-sigma detection threshold of 1.52 mJy 
(or 1.59 mJy when using the median).  Taking a safety margin, we can state that the sensitivity 
of the blind survey is about 1.7 mJy for one 160-s integration with the backend installed in July 2006.

\begin{figure}
\centering
\includegraphics[width=8.5 true cm]{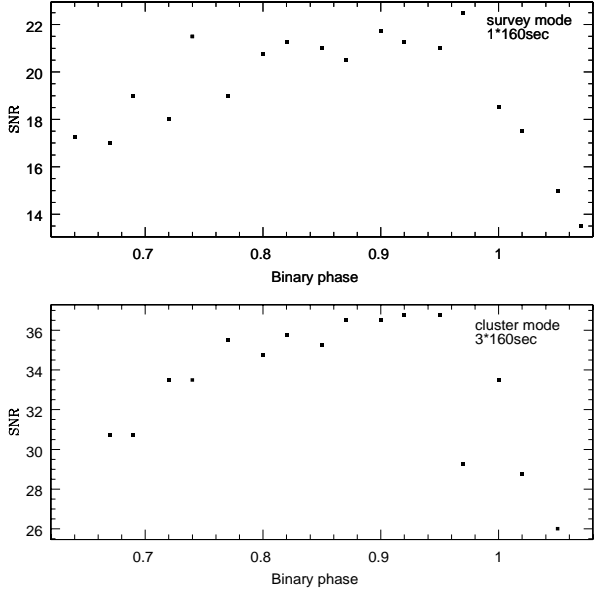}
\caption{Signal-to-noise ratio for observations of \object{PSR B1744-24A}
with third setup.
Top: single 160-s integrations and reduction in ``survey mode". 
Bottom: three 160-s integrations and reduction in ``cluster mode".
}
\label{figsnr1744}
\end{figure}

\subsection {Performance validation for ``cluster mode"}
\label{perfhough}

We have validated the strategy for processing consecutive observations 
of an accelerated pulsar by applying it 
to the binary pulsar \object{PSR B1744-24A} in the globular cluster \object{Terzan 5}. 
This pulsar is strong enough (when not in eclipse) to be easily detected 
and is in a close enough binary system to display relatively large variations in
the period and period derivative within typical observing times (a few 160-s observations). 
We carried out 18 consecutive 160-s observations on 16 March 2007. 

The 18 observations were then processed in 16 groups of three observations, 
following the procedure described in Sect. \ref{proc_n_obs}. The detection SNR values obtained 
in these analyses are shown in Fig.~\ref{figsnr1744}.
They are perfectly consistent with the coherent addition of the SNR values 
obtained by treating all integration separately, except for one point at binary phase 0.98 
for which the assumption that the pulsar period varies linearly does 
not hold sufficiently over three consecutive observations.

In our search for new pulsars, we also observed globular clusters with known pulsars, 
which allowed us to check the processing in ``cluster mode" (the same validation can also be performed 
on field pulsars, but at the cost of some extra observing time). Table \ref{tabletest} presents a summary of the results
that we now explain in some more details. As an example, \object{PSR J1807-2459A} in the cluster \object{NGC 6544} 
has an expected flux density of 1.1 mJy at 1400 MHz, thus would yield an SNR that is quite insufficient 
for detection using only one 160-s integration.  Processing of three consecutive observations yields an SNR of 8.4, 
allowing detection in a wide search. Similar detections were obtained for several other pulsars. 
For \object{PSR B1821-24} in \object{NGC 6626}, the flux density is 0.18 mJy in ATNF, but 1.1 mJy in Lyne et al. (\cite{lyne87}). 
We obtained an SNR of 8 for four successive 160-s observations, which seems to favor the second value. 
For \object{PSR B1620-26} in \object{NGC 6121}, with an expected flux density of 1.6 mJy, 
we obtained an SNR of 19 for four 160-s observations. For the field eclipsing 
pulsar \object{PSR J2051-0827}, the expected flux density is quite variable and is 
reported at 2.8 mJy in ATNF, but only 1.1 mJy in Stappers et al. (\cite{stappers}). 
We obtained an SNR of 9 for four 160-s observations, which is consistent with the second value. 
Overall, we can expect to obtain a 8-sigma detection threshold at about 1 mJy or 
slightly below with four consecutive observations.

\begin{figure}
\centering
\includegraphics[width=8.5 true cm]{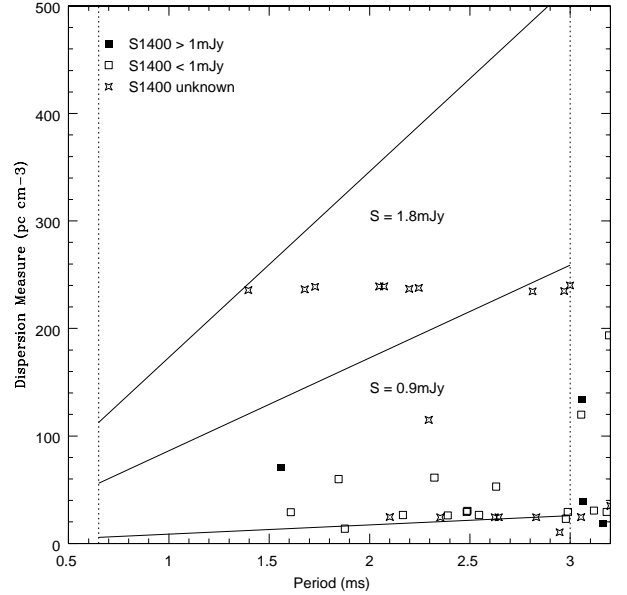}
\caption{Dependence of the sensitivity on dispersion measure.
The sensitivity decreases progressively toward higher dispersion measures. 
The diagonal lines correspond to a dispersion representing 0.1, 1, and 2 windows per frequency channel, respectively.
The individual points are known millisecond pulsars with different symbols depending
on their flux density at 1400 MHz.}
\label{figsnrlimit}
\end{figure}

\subsection {Dependence of the actual sensitivity on dispersion measure}

The sensitivities quoted above are nominal sensitivities, and the actual ones will depend on
the dispersion measure at the distance of the pulsar. This is true for any processing method.

In our method, the power spectrum is computed over time windows of a small fraction of 
the tested period and typically uses 128 frequency channels.
Because of the 1-bit sampling the spectrum is normalized and we rely on the variation in the power (with time across the windows 
and with frequency across the channels) due to the dispersion of the pulse to detect a pulsar.
The effect of the dispersion measure on the sensitivity is shown in Fig.~\ref{figsnrlimit}. 
Starting from the bottom, the first diagonal line is the limit below which 
we can hardly detect pulsars because the dispersion is too low (less than 0.1 window per frequency channel),
and the signal is spread equally in all frequency channels. 
As the dispersion increases, the signal for a given window in time is spread over fewer frequency channels, 
and contrast appears between channels with and without signal.
The sensitivity is optimal (i.e. 0.9 mJy for four integrations of 160 s and reduction in ``cluster mode") 
in the region between the first and second diagonal line, i.e., between 0.1 and 1 window per frequency channel. 
Above the second diagonal line, the sensitivity decreases by a factor 2 because 
the dispersion is between one and two windows per frequency channel. Above the top diagonal line, 
the sensitivity loss is such that we simply did not search that part of the parameter space during processing.
These are not strict limits, however, at neither the low nor the high end.
For example, the signal of the pulsar is spread over 20 channels out of 128 for a dispersion of 0.05 window per frequency channel, 
so there is still some contrast in the power spectrum between a window with no signal
and a window with a signal in some part of the band. For high dispersion values, the
pulse is spread over several windows so the fraction of the signal in
each window is reduced, and furthermore the contrast in the power spectrum of a given window
is also reduced, because the signal will appear in different channels for different pulses.

There are other practical reasons for limiting the search to the interval between 0.1 and 2 
windows per frequency channel. Below 0.1, pulsars with such low dispersion must 
be nearby and should already have been detected, and above 2, pulsars with such high dispersion measure are
more likely to be distant, thus too faint given the decreased sensitivity.

\begin{figure}
\includegraphics[width=9.0cm]{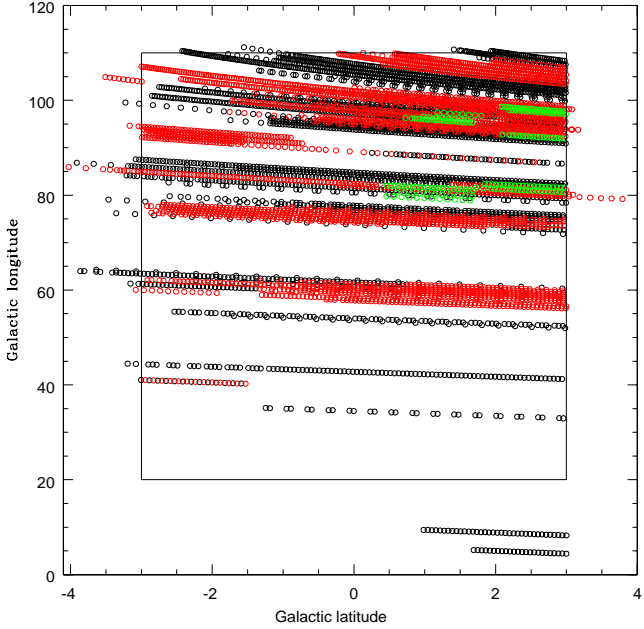}
 \caption{Map of the observed points in the blind survey in the Galactic plane.
The green, red, and black open circles correspond
to observations with the first, second, and third setups, respectively.
The size of the symbols is constant and not representative of the
radiotelescope beam, which is elongated in declination and 
increases with declination (thus roughly with longitude). 
The inner rectangle encloses the region of the nominal survey.
Regions below $l = 70^{\circ}$ are poorly
covered because of oversubscription of the radiotelescope time at lower
longitudes.
}
\label{figmap}
\end{figure}

%__________________________________________________________________
\section{The searches}
\label{searches}

Our initial goal was to do a blind survey in the Galactic plane. But after
two years of unsuccessful searching, 
we estimated that the probability of return per hour of observing time was probably 
higher in targeted searches than in a blind survey,
so we decided to try them as well.

There are several possible types of targeted searches for millisecond pulsars.
The most successful by far is to point at Galactic globular clusters. This
is where most recent searches for such objects have been done
(e.g. Possenti et al \cite{possenti}, Ransom et al \cite{ransom}, Freire et al \cite{freire}).
Another option is to select radiosources with no optical counterpart in catalogs of
radiosources, as done by several groups (Crawford
et al. \cite{crawford}, Han et al. \cite{han}), and with success by Navarro et al (\cite{navarro}).
Pulsars can be distinguished from other radiosources in blind surveys by their
steep spectral index and by a high degree of polarization, usually above 5\%.
Millisecond pulsars tend to show the same, sometimes
high, degree of linear polarization as ordinary pulsars
(Yan et al. \cite{yan}).  

\begin{table}
\caption{List of pulsar candidates with SNR$>6.5$. 
}
\centering
\label{tablecand}
\begin{tabular}{llrrr}
\hline
Name&RA(B1950)&Dec(B1950)&$P$&N\\
    &         &          &ms & \\
\hline
4091907&20 00 25.36 &+25 10 0 &0.876&10\\
2541105&20 10 18.22 &+37 16 0 &0.885& 4\\
2271416&20 18 00.63 &+37 16 0 &1.071& 4\\
4830818&20 29 16.4  &+43 52 0 &1.047& 7\\
3690234&20 29 22.06 &+38 22 0 &1.171&11\\
5191628&20 30 44.2  &+40 56 0 &0.723& 3\\
5141841&20 33 54.2  &+40 56 0 &4.347& 3\\
4481402&20 35 56.31 &+41 40 0 &0.892& 4\\
3760145&20 41 19.79 &+38 44 0 &1.742&10\\
4942356&20 41 51.00 &+43 52 0 &0.997& 5\\
4780947&20 48 30.47 &+43 52 0 &3.579& 8\\
4561319&20 54 08.7  &+41 40 0 &2.628& 4\\
       &            &         &0.920&  \\
4721049&21 13 53.4  &+43 08 0 &2.009& 4\\
3630610&21 36 02.95 &+51 56 0 &0.954& 4\\
3680335&21 41 02.86 &+54 52 0 &1.738& 4\\
3161822&22 02 25.0  &+58 54 0 &2.747& 4\\
2791002&22 07 40.95 &+55 36 0 &1.720& 4\\
3072105&22 08 50.2  &+58 32 0 &1.783& 4\\
4251900&22 10 55.6  &+54 08 0 &0.675&10\\
4012312&22 16 51.3  &+54 08 0 &0.952& 7\\
3092112&22 32 51.15 &+58 32 0 &1.425& 4\\
4032330&22 42 30.55 &+61 28 0 &1.248& 7\\
2210226&22 49 20.78 &+60 22 0 &2.108& 4\\
2140142&20 15 12.55 &+36 32 0 &2.489& \tablefootmark{a}\\
3451149&22 09 12.42 &+54 30 0 &1.492& \tablefootmark{a}\\
\hline
\end{tabular}
\tablefoot{
\tablefoottext{a}{not reobserved for lack of observing time}
}
\end{table}

\begin{table*}
 \centering
  \caption{List of the observed globular clusters}
     \label{tablegc}
  \begin{tabular}{lrrrr|lrrrr}
  \hline
Cluster&$DM$&Np& date &N&Cluster&$DM$&Np& date &N\\
       &pc cm$^{-3}$&  &yyyymmdd& &       &pc cm$^{-3}$&  &yyyymmdd& \\
  \hline
\object{NGC 5986}&169-171&0&20070213& 4&        &   &&20070815&18\\
\object{NGC 6093}&108-109&0&20070312& 4&        &   &&20070816&18\\
\object{NGC 6121}& 63&20&20070213& 4\object{&NGC 6553}&237-350&4&20070313& 4\\
\object{NGC 6171}&83-86&4&20061119& 3&        &   &&20071211& 8\\
        &   &&20070213& 4&\object{NGC 6624}& 87&0&20070427& 4\\
        &   &&20070426& 4&\object{NGC 6626}&121&5&20070211& 4\\
        &   &&20071005&10&\object{NGC 6712}&235-333&3&20061114& 3\\
        &   &&20071201&10&        &   &&20070313& 4\\
\object{NGC 6218}&70-75&7&20061119& 3&        &   &&20071221&18\\
        &   &&20070213& 4&\object{NGC 6717}&163-193&2&20061114& 3\\
\object{NGC 6254}&73-81&8&20061119& 3&        &   &&20070923&18\\
        &   &&20070312& 4&\object{NGC 6760}&208-310&1&20061203& 4\\
        &   &&20070426& 4&\object{NGC 6809}&103-113&6&20061203& 4\\
        &   &&20070624&18&        &   &&20070117& 4\\
        &   &&20070718&18&        &   &&20070504&18\\
        &   &&20071101&10&        &   &&20071121&10\\
\object{NGC 6256}&418-554&0&20071201&10&\object{2MASSGC01}&144-292&13&20070210& 4\\
\object{NGC 6266}&115&3&20070313& 4&         &   &&20070719&18\\
\object{NGC 6304}&224-312&4&20070119& 4&\object{2MASSGC02}&127-175&7&20070211& 4\\
        &   &&20070210& 4&         &   &&20070922&10\\
        &   &&20070815&18&\object{FSR1767} &61-110&25&20070922& 8\\
\object{NGC 6342}&103&0&20070426& 4&\object{Glimpse-CO2}&294-560&5&20080919& 4\\
\object{NGC 6366}&69-85&13&20070119& 4&           &   &&20081122& 5\\
        &   &&20070516& 4&\object{IC 1276} &176-251&5&20070427& 4\\
\object{NGC 6440}&224&0&20070427& 4&        &       & &20071113&18\\
\object{NGC 6441}&366-451&0&20070427& 4&\object{Terzan10}&268-413&5&20071211& 8\\
\object{NGC 6544}&134&16&20061114& 3&\object{Terzan12}&197-332&7&20061114& 3\\
\object{NGC 6540}&212-315&13&20070119& 4&        &   &&20070211& 4\\
        &   &&20070516&18&        &   &&20071012&18\\
\hline
\end{tabular}
\end{table*}

\subsection{The blind survey}
\label{blind}

The blind survey was conducted in the Galactic plane, where most pulsars are likely to be: 
latitude ($b$) smaller than 3$^{\circ}$, longitude ($l$) between 20$^{\circ}$ and 110$^{\circ}$.
One could argue that in setting a limit in $b$ rather than in height above (or below) 
the Galactic plane, we miss nearby pulsars that are at high $b$ {\it and} in the Galactic plane.
However, as shown below (Sect.~\ref{resultsimu}), the gain in detectable pulsars would be largely 
offset by the cost in telescope time.

A typical observing run was done at a set declination and 
lasted 20 to 120 minutes (depending on the declination and the alloted time),
shifting the telescope (in fact the receiver, since this is a transit instrument) 
by one beam in right ascension after each short integration. 
A better observing strategy was to shift by multiples of one beam
and return to that declination at later dates to fill the gaps. The selected declinations were
also separated by one beam, so that all pointings were separated by at least one beam in either
direction. 

The total amount of telescope time attributed to this project over the years was 549.3 hours, at
which point the time allocation committee wisely decided to cut its losses.
We observed 186 points in the sky with the first setup (sensitivity $\simeq$ 3.5 mJy for integrations of 240 s), 2829 points with the
second setup (sensitivity $\simeq$ 2.2 mJy for integrations of 200 s) and 2542 points with the last setup (sensitivity
$\simeq$ 1.7 mJy for integrations of 160 s). 
Seventy points were observed several times, leading to a total of 5487 different points 
observed out of a total of 22057, that is, a coverage of 25\% of the planned survey.
The actual sensitivity was different in some of the pointings.
In the second semester of 2004, the sensitivity progressively degraded from 2.2 to 2.7 mJy due to an unnoticed desynchronization in the
acquisition. A bug in the treatment software in late December 2004 lowered the sensitivity by about half for 57 points.
We attributed a slightly higher sensitivity (by an arbitrary amount of 20\%) to the 70 points observed repeatedly, 
because of the better chance of detecting pulses as the effective flux may vary and as the fixed set of tested values in the parameter space 
may better match the actual parameters of the pulsar.  

A map of the survey region in Galactic coordinates with the covered region shown as circles
(colors depending on setup) is presented in Fig.~\ref{figmap}.
The figure shows that a few points were observed outside the region of the survey,
either by mistake, or by opportunity (to save repointing the telescope). 
The figure also shows that regions below $l =  70^{\circ}$ are poorly
sampled, because of oversubscription of the telescope time 
by a factor greater than 2 at the corresponding right ascensions.  
While millisecond pulsars are likely to be more frequent at
lower Galactic longitudes (closer to the Galactic center), their {\it a priori} detectability is
strongly reduced by pulse smearing at longitudes below 50$^{\circ}$
(see Sect.~\ref{mspopulation} and Fig.~\ref{figsimu}). 
However, the drawback of higher declinations (and thus higher longitudes) is that the antenna aperture
efficiency is progressively reduced, with a loss of sensitivity
of 15\% at $l \simeq 85^{\circ}$ and 28\% at $l \simeq 110^{\circ}$ (Fouqu\'e et al. \cite{fouque}).
About half of our pointings suffer a loss of sensitivity of 20\% or more because of that effect.

During the survey we identified 25 pulsar candidates with an SNR$\simeq$7. 
They are listed in Table~\ref{tablecand}, together with their coordinates (1950 equinox), probable period
(in ms) and the number of 160-s integrations in the reobservations. Note that the last two candidates were not
reobserved for lack of observing time.
Their reobservation and analysis in ``cluster mode"
(see Sect.~\ref{proc_n_obs}) did not enable us to confirm them.
Some of these presumably false detections can be explained by interferences from
radars at civilian and military airports near Paris.

\subsection{Globular clusters}
\label{gc}

The central regions of globular clusters are favorite targets when searching
for millisecond pulsars, because they are regions of very high stellar
density. In such regions, multiple encounters between stars favor the formation
of binary stars (Bailyn \cite{bailyn}), and, when one of the stars is a pulsar, favors the
acceleration of the pulsar to periods of about a millisecond (pulsar recycling). 
Furthermore, a large number of binary stars in globular clusters may in fact 
be primordial binaries (Hut et al. \cite{hut}), thus increasing the probability of finding
recycled pulsars in globular clusters. 

We selected globular clusters as targets using three criteria:
\begin{itemize}
  \item Avoid the zone accessible to the radiotelescope in Arecibo, which, with
its large collecting area and its new ALFA receiver, can conduct much
more sensitive searches than us.
  \item Aim at nearby globular clusters, with high stellar density, and 
where no millisecond pulsars have been detected so far.
  \item Select preferentially those clusters where the dispersion measure is
highest, and hence where pulsars are difficult to detect with some of
the search techniques used in past surveys (see below).
\end{itemize}

To test the feasability of our project, we selected in the ATNF the 32 pulsars discovered
in globular clusters as of April 2006 with a published flux density. 
If the flux density was given at 400 MHz, it was converted to 1400 MHz assuming a spectral index
of -1.8.  We then calculated the number Np of these 32 pulsars that would be detected if they were all
in each of the different Galactic globular clusters. 
We selected those clusters in which we have a reasonable chance of detected millisecond
pulsars taking Np, the sensitivity of our search, and the value of the 
dispersion measure at the distance of the cluster into account.  Other clusters were added 
in the course of the search to test the sensitivity for 
reduction in ``cluster mode" (see Sect.~\ref{perfhough}) 
or to optimize the use of the available telescope time.

We obtained 28.33 hours (+ 3 of discretionary director's time) of telescope time for this project.
The list of observed globular clusters and their characteristics
are given in Table~\ref{tablegc}. The successive columns are the name of the cluster,
the range in dispersion measure (from the model of Cordes \& Lazio \cite{cordes02}), or the dispersion measure
(from the ATNF catalog) 
if millisecond pulsars have been detected in the cluster, in cm$^{-3}$pc, 
the expected number Np of pulsars with a flux density above 1 mJy
(assuming that all 32 know millisecond pulsars are in that cluster), 
the date of observation, and the number of pointings.

Candidates were detected in the clusters \object{NGC 6171}, \object{NGC 6254} and \object{NGC 6809}, which were reobserved repeatedly.
No new millisecond pulsar was detected.

\subsection{Bright polarized point-like radiosources with steep spectra}

In the second targeted search, the candidate pulsars were selected from the
catalog of steep spectral index radiosources published by De Breuck et
al. (\cite{debreuck}). While their catalog was constructed with the aim of obtaining
a large sample of high-redshift radiosources, because spectral index tends
to increase with the redshift of optically identified sources, 
a large negative spectral index is also a property of millisecond pulsars.

We extracted the information on polarization and on
source size from the NVSS\footnote{http://www.cv.nrao.edu/nvss/} 
catalog of radiosources.  From this complete list, we extracted
the sources that were unresolved and polarized ($\geq$ 5\%). 
These criteria give
26 sources, two of which are in fact known pulsars (\object{PSR J0630-2834} and
\object{PSR J2313+4253}). We removed them
from the list and checked that the 24 other sources are not in any database
(Simbad, NED, Hyperleda). The final list is given in Table~\ref{tablessr}, together with
their 1400-MHz flux density in mJy, spectral index between 1400 and 326 MHz, 
percent polarization, the date of our observation, and the number
of 160-s integrations. More information on the sources can be found
in De Breuck et al. (\cite{debreuck}). 

We obtained 18.67 hours of telescope time for this project and observed all
the sources in the list. None of them turned out to be a pulsar.

\begin{table}
 \centering
 \begin{minipage}{\columnwidth}
\caption{List of observed bright polarized unidentified radiosources with steep spectrum}
\label{tablessr}
\begin{tabular}{lrlrcr}
\hline
Name          & S1400&$\alpha$&\% pol.&date&N\\
              & mJy  &        &       &yyyymmdd& \\
\hline
WN J0751+3300 & 10.7 &-1.47& 5&20080107 & 18\\
WN J0829+3834 & 10.9 &-1.49& 5&20080202 & 18\\
TN J0856-1510 & 68.5 &-1.41& 12&20080304 & 18\\
WN J0928+6003 & 10.4 &-1.63& 7&20080402 & 18\\
WN J1101+3520 & 14.9 &-1.32& 5&20080204 & 18\\
WN J1111+3311 & 12.3 &-1.32& 6&20080319 & 18\\
WN J1152+3732 & 14.7 &-2.18& 5&20080110 & 18\\
WN J1258+3212 & 18.3 &-1.52& 8&20080306 & 18\\
WN J1300+5311 & 18.1 &-1.44& 5&20080214 & 18\\
WN J1330+6505 &  9.5 &-1.34& 5&20080413 &  3\\
WN J1337+3401 & 13.2 &-1.39& 7&20080116 & 12\\
WN J1359+7446 & 11.2 &-1.92& 5&20080413 & 4\\
WN J1400+4348 & 18.2 &-1.49& 6&20080205 & 4\\
WN J1421+3103 & 11.6 &-1.43& 5&20080413 & 3\\
WN J1435+3523 & 20.8 &-1.35& 7&20080320 & 12\\
TN J1515-2651 & 56.7 &-1.34& 9&20080218 & 18\\
WN J1606+4142 & 10.9 &-1.31& 10&20080407 & 18\\
WN J1723+5822 & 10.1 &-1.46& 11&20080111 & 8\\
WN J1734+3606 &  7.1 &-1.31& 14&20080117 & 8\\
WN J1739+5309 &  9.8 &-1.32& 5&20080111 &  8\\
WN J1819+6213 & 11.6 &-1.45& 7&20080123 & 12\\
WN J1923+6047 & 11.8 &-1.30& 8&20080419 & 18\\
TN J2028-1934 & 131.4 &-1.34& 7&20080302 & 15\\
WN J2221+3800 & 13.2 &-1.32& 5&20080422 & 18\\
\hline
\end{tabular}
\end{minipage}
\end{table}

\subsection{Faint polarized point-like radiosources}

In the third search, we extracted the candidate pulsars from the NVSS catalog.
This catalog contained 97 of the 1775 pulsars in
the ATNF catalog at the time of the project (early 2008). 
From the NVSS catalog, we first extracted the sources
with 1400-MHz flux densities between 2 and 10 mJy.  
We did not try to determine their spectral index, as in the previous set of
targets, because there is no
hope of finding these faint sources in catalogs at other frequencies.

We then selected those sources that were polarized 
(polarized flux density $>$ 20\% of total flux density)
and with SNR in total and polarized flux density above 3. This
produced a list of 2105 sources. In this list, we found six pulsars
(\object{PSR J0358+5413}, \object{PSR J0538+2817}, \object{PSR J0814+7429}, \object{PSR J0922+0638}, \object{PSR J1745-3040}, \object{PSR J1840+5640}),
none of which has a very short period. But one should note that none of the 45
known millisecond pulsars with periods less than 3.3 ms has 1400-MHz flux 
densities above 2 mJy, except \object{PSR B1937+21}, so they cannot be in that catalog.
We then further restricted the sample to the Galactic plane ($|b| < 12^{\circ}$),
also eliminating the region $75^{\circ}<l<165^{\circ}$ and $|b|<5.5^{\circ}$ 
(recently surveyed by Hessels et al. \cite{hessels08}),
the region of our own blind survey, and the right ascensions heavily
oversubscribed at Nan\c cay (17h00-19h40).  We removed all sources identified
in NED, SIMBAD, and 2MASX. 
We also carefully examined the postage-stamp images of
all the sources and removed those that looked like noise, as well as close
double sources (which are likely to be double-lobed radiogalaxies). This brought the
total number of targets to 92.  Some ambiguous cases (multiple sources, not
quite point-like sources) were left in the sample, since at this low flux level
it is sometimes difficult to distinguish real features from noise. 

We were given 21.67 hours of telescope time for this project, which enabled us 
to observe 65 of them (with one source observed twice).
They are listed in Table~\ref{tablefs}, which gives the
NVSS name, the flux density at 1400 MHz in mJy, and the percent polarization.  
This last targeted search did not produce any new millisecond pulsar, which seems to confirm
that most radiosources with ultrahigh polarization are radio-loud active galactic nuclei
(Shi et al. \cite{shi}).  

The conclusion of the two targeted searches in catalogs of radiosources
is that the luminosity function of millisecond pulsars is simply too faint
to allow a meaningful number of them to be found in large surveys like NVSS
or FIRST. The small number of sources observed does not allow us to be more
quantitative.
 
%__________________________________________________________________
\section{The Galactic population of millisecond pulsars}
\label{mspopulation}

We now examine whether the null result of our blind survey is statistically compatible with 
the Galactic distribution of millisecond pulsars as we know it.  The method is the following. We adopt a
model for the distribution of millisecond pulsars in space, rotation period, and luminosity,
produce a large number of Galactic populations of synthetic millisecond pulsars,
and, for each realization of the simulation, we compute how many synthetic pulsars 
should be detected in our blind survey, by comparing the properties of each synthetic pulsar
(position, rotation period, flux density) with the coordinates and actual sensitivity of each of our 5487 pointings,
taking the effect of the dispersion measure on the detectability into account in each case. 
The number of detected pulsars should be zero in most realizations, otherwise 
our model overestimates the true number of millisecond pulsars.

\subsection{Model for the Galactic population of pulsars}
\label{model}

Several models have been proposed over the years to characterize the population
of pulsars in the Galaxy, either ordinary
(Lyne et al. \cite{lyne85}; Yusifov and K\"u\c c\"uk, \cite{yusifov04};
Lorimer et al. \cite{lorimer06}; Faucher-Gigu\`ere \& Kaspi \cite{faucher06}),
millisecond
(Lorimer et al. \cite{lorimer96}; Cordes \& Chernoff, \cite{cordes97}; Story et al. \cite{story}),
or both (Lyne et al. \cite{lyne98}; Faucher-Gigu\`ere \& Loeb, \cite{faucher10}).

The population of millisecond pulsars is too small to allow
for a precise modeling of their spatial distribution.
However, they are thought to have larger radial and vertical 
scale lengths than those of ordinary pulsars, because they are
likely to have lost the memory of their birth location due to a high initial velocity.

\begin{figure}
\includegraphics[width=8.5cm]{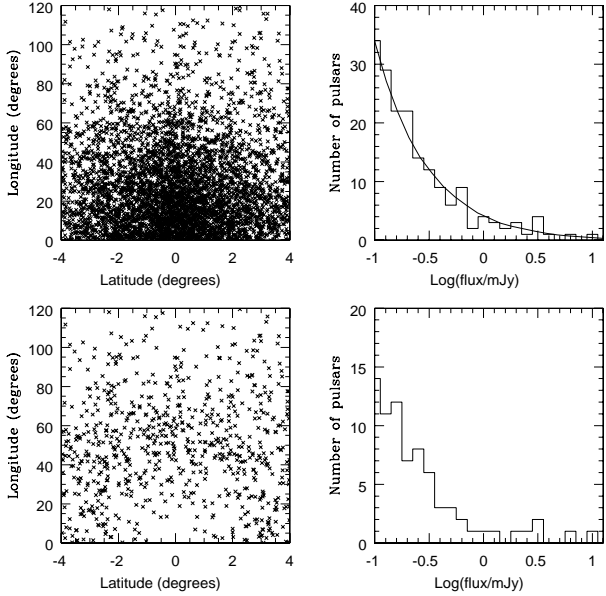}
\caption{One simulation of the population of 30~000 millisecond pulsars
in the Galaxy. {\it Top:} Distribution of all (4589) short-period pulsars in the region of 
the blind survey and corresponding histogram of flux densities; the solid curve
is the mean histogram from 1000 simulations. 
{\it Bottom:} distribution of the (730) pulsars remaining
after eliminating the undetectable ones (because of pulse smearing
at high dispersion measure) and corresponding histogram of flux densities.
Both histograms are limited to flux densities higher than 0.1 mJy for clarity. 
}
\label{figsimu}
\end{figure}

We adopt the spatial distribution used by Faucher-Gigu\`ere \& Loeb (\cite{faucher10}).
The radial density in the Galactic plane is $ \rho(r) \propto exp(-r^2/2\sigma^2) $,
where $r$ is the radial distance from the Galactic center and $\sigma$ =  5 kpc. 
For ordinary pulsars, $\sigma$ = 3.6 kpc (Yusifov and K\"u\c c\"uk, \cite{yusifov04}).
Our simulations and others (Lorimer et al. \cite{lorimer96}) have shown that 
the exact radial dependence is in fact not important, because
most millisecond pulsars are faint, and
the heavily populated region toward the Galactic center
is ``obscured" by a high dispersion measure (see Fig.~\ref{figsimu}).
Most detectable millisecond pulsars are thus within 2 or 3 kpc
of the Sun (which in this paper is assumed to be located 8 kpc from the Galactic center).
The distribution perpendicular to the Galactic plane is a simple
exponential~: $ \rho(z)  \propto exp(-|z|/h) $,
where $z$ is the distance from the Galactic plane and the scale height $h$ = 500 pc
(Cordes \& Chernoff \cite{cordes97}, Story et al. \cite{story}).
This value is at the upper end of those estimated for ordinary pulsars
(330-600 pc).
Adopting a higher value for the scale height slightly reduces the number of
pulsars at low Galactic latitudes, and hence the probability of detection in
our blind survey. We do not take into account a possible increase in $h$ with $r$ (Yusifov \cite{yusifov}),
which would reduce the number of pulsars in the Galactic plane even more. 

The total number of millisecond pulsars is highly uncertain. 
Lorimer et al. (\cite{lorimer95}) estimate that there are 40~000 millisecond 
and low-mass binary pulsars
above $L_{min}$ = 2.5 mJykpc$^2$ at 436 MHz.
Lyne et al. (\cite{lyne98}) estimate that there are
30~000 millisecond pulsars above $L_{min}$ = 1 mJykpc$^2$ at 400 MHz.
Desvignes (\cite{desvignes}2009) finds 46~000 millisecond pulsars above  $L_{min}$ = 0.1 mJykpc$^2$
at 1400MHz using model C of Lorimer et al. (\cite{lorimer06}), which amounts to 23~000
above 0.2 mJykpc$^2$ with our adopted luminosity function.
We adopted a total number of 30~000 millisecond pulsars above
$L_{min}$ = 0.2 mJykpc$^2$ at 1400 MHz, which is a happy medium
between the above estimates, using a spectral slope of -1.8
to convert $L_{min}$ to the right frequency. 

The observed distribution of millisecond pulsar rotation periods $P$
is a power law of index +3 (Hessels et al. \cite{hessels07}). However, 
the paucity of fast-spinning pulsars is the result of selection effects
(see Sect.~\ref{intro} and Lorimer et al. \cite{lorimer96}).
Following Lorimer et al. (1996), we adopt a distribution of rotation periods
that is uniform in log($P$). The lower limit of 0.65ms is the lowest
theoretically possible one (Haensel et al. \cite{haensel}), and the upper limit 
reflects an observed cutoff. Cordes \& Chernoff (\cite{cordes97}) assume a period
distribution in the form $P^{-1}$.
For the rotation periods of ordinary pulsars, 
a log-normal distribution is generally assumed (Lyne et al. \cite{lyne85}, Lorimer et al. \cite{lorimer06}).

The distribution of luminosities is also modeled by a power law, 
down to the lowest luminosity (Lorimer et al. \cite{lorimer07}, 
Hessels et al. \cite{hessels07}, Bagchi \& Lorimer \cite{bagchi}):
$ \rho(L) = L^{-n} $ if $L > L_{min}$, and 0 otherwise, with $n$ around unity.
We adopted $n$ = 1. 
The luminosity distribution of ordinary pulsars has been
modeled by one or two power laws (Lyne et al. \cite{lyne98}, Guseinov et al. \cite{guseinov}).
Models where the luminosity distribution depends
on the pulsar period and its derivative have also been considered,
for both millisecond and ordinary pulsars (Story et al. \cite{story}).
The flux density was computed by multiplying the luminosity by the square of
the distance to the synthetic pulsar.

Finally, we take possible pulse smearing by the interstellar
medium into account, using the Galactic free electron model 
NE2001\footnote{http://rsd-www.nrl.navy.mil/7213/lazio/ne\_model} 
of Cordes \& Lazio (\cite{cordes02}) to estimate the dispersion measure ($DM$) at the
distance of the synthetic pulsars.  If $DM >  86.38P$, where $P$ is the period
in ms and $DM$ is in cm$^{-3}$pc, the sensitivity is reduced by half.  
If $DM > 172.76P$, the sensitivity is reduced by a factor 3 or more, and we consider that
the pulsar is undetectable (see Fig.~\ref{figsnrlimit}).
On the other hand, we ignore the effects on the flux density 
of diffractive interstellar scintillation, which can 
make the measured flux density vary by a large factor, 
as shown for \object{PSR B1937+21} in Fig.~\ref{figsnr1937}.

\subsection{Results of the simulations}
\label{resultsimu}

We did many simulations of the distribution and detectability of the
pulsar population in the region of our blind survey, using the above model and parameters.
Out of a population of 30~000 synthetic pulsars in the Galaxy, only about
4600 short-period millisecond pulsars ($P \leq 3$ms) fall in the region of the survey extended to 
$0^{\circ} < l < 120^{\circ}$ and $|b| < 4^{\circ}$ for comparison with Fig.~\ref{figmap}.
There are about 5700 long-period millisecond pulsars ($P > 3$ms) in that region, which thus contains a
bit more than a third of the total population of synthetic pulsars.
We show the result of one simulation in Fig.~\ref{figsimu}. 
The distribution of the short-period pulsars is shown in the top panel, together
with the histogram of flux densities above 0.1 mJy. 
The smooth curve is the mean histogram for a thousand simulations.
The pulsars are concentrated in the region nearest to the Galactic center.

However, when considering the effect of high $DM$, 
that is, eliminating all the pulsars above the top diagonal line
of Fig.~\ref{figsnrlimit}, the overall
population of detectable pulsars in the region of the survey is reduced by a factor 6. 
This is shown in the bottom panel of Fig.~\ref{figsimu}, where the region near the Galactic center
is no longer heavily populated, and the region $40^{\circ} < l < 80^{\circ}$ 
becomes the most populated. 
This depopulation of the detectable pulsars is not so critical if one only considers the
pulsars with a high flux density, say above 1 mJy. This population is reduced from 17 to 9, that
is, by about half, because the depopulation affects mainly pulsars with a high $DM$, 
which are necessarily more distant, thus fainter and more difficult to detect. 
  
The number of detectable pulsars for a given sensitivity cannot be read directly from
the histogram of the lower panel of Fig.~\ref{figsimu}, because the actual sensitivity
will depend on the $DM$ at the distance of the pulsar.
Taking the example of our project, we find that 4.2 pulsars in the extended
region of our survey are detectable for the nominal sensitivity of 1.7 mJy of our last setup
(taking the effect of large $DM$ into account), and 7.7 pulsars are detectable for
a sensitivity of 0.9 mJy (which is achieved with four integrations of 160 s
and reduction in ``cluster mode"). This means that,
if we want to double our chances of
detecting a pulsar, we have to increase our sensitivity twofold. This can be achieved by a larger bandwidth, 
but, if keeping the equipment unchanged, 
it would require increasing the telescope time fourfold, which is not an economical strategy.
It is more efficient to extend the region of
the survey rather than increase the sensitivity 
(provided the added region is as rich in potential pulsars as the initial region). 
Only at much higher sensitivities does the
number of synthetic pulsars go up significantly with increased sensitivity (see histograms of Fig.~\ref{figsimu}).

Now, to estimate whether a synthetic pulsar is detected in any
one of our pointings, we model the beam pattern of the radiotelescope
by a $sinc^2$ function in both coordinates (modeling by a Gaussian is only acceptable
within the half-power beamwidth), and compute the sensitivity correction
at that pointing as $\Delta s$ = $sinc^2(d\alpha)sinc^2(d\delta)$,
where $d\alpha$ and $d\delta$ are the angular distances of the synthetic pulsar
to the center of the beam in right ascension and declination, respectively.  
The half-power beamwidth is 3.6'x22' at 1400 MHz and zero declination, and
there is an additional correction in declination above 22.7$^{\circ}$. 
The beamwidth is about a factor 1.32 larger in declination at our highest declination. 
We also consider the loss of gain due to variable aperture efficiency and atmospheric absorption, 
using the equation given by Fouqu\'e et al. (1990, their equation 1).

The results of 200 000 simulations are given in Table~\ref{tableproba}, where
$N_p$ is the number of detected synthetic pulsars, 
$P_s$ the probability of detecting $N_p$ pulsars in our survey of 5487
pointings, $P_{1.7}$ the probability of detecting $N_p$ pulsars
if we had completed the whole survey with a nominal sensitivity of
1.7 mJy from beginning to end, and $P_{0.9}$ the probability of detecting $N_p$ pulsars
if we had surveyed the more densely populated region $|b| < 2^{\circ}$ and $45.0^{\circ} < l < 68.0^{\circ}$
(3761 pointings) with a higher sensitivity of 0.9 mJy (by doing four 160-s integrations at each pointing 
and reducing the data in ``cluster mode"). 

In summary, we had one chance in four
of detecting one or more short-period pulsars in our incomplete survey.
Were we to redo the complete survey with the sensitivity reached in the
last setup, we would have a 79\% chance of success, but even
so the harvest would be far from spectacular. And this ignores the effect of
interstellar scintillation, which reduces our chances of success by an unknown amount.
While the third option of surveying a smaller region with a higher sensitivity
may {\it a priori} seem attractive, it has a low return 
and is in practice unfeasible at a transit instrument like the Nan\c cay
radiotelescope, because it would monopolize the right-ascension slot 19h-20h for
about two years.

As mentioned in Sect.~\ref{blind}, we miss nearby high-latitude millisecond pulsars 
when limiting the blind survey to low latitudes. If we set no limit in $b$ (there is
of course no way of limiting the survey in linear height above the Galactic plane),
we find that there are about 6510 (rather than 4600) short-period millisecond pulsars
in the region of our survey, 1980 (rather than 730) detectable ones, and 29
(rather than 9) above 1 mJy.
However, we would have to scan about half the sky visible at Nan\c cay to complete the
survey, so this is another unrealistic option.

\begin{table}
 \centering
 \begin{minipage}{\columnwidth}
  \caption{Probability of detecting short-period pulsars
}
     \label{tableproba}
  \begin{tabular}{r|lll}
  \hline
$N_p$&$Ps$(\%)&$P_{1.7}$(\%)&$P_{0.9}$(\%)\\
  \hline
0&0.763&0.213&0.697\\
1&0.151&0.223&0.174\\
2&0.055&0.184&0.073\\
3&0.020&0.145&0.035\\
4&0.006&0.099&0.014\\
5&0.003&0.059&0.004\\
6&0.0006&0.038&0.0014\\
7&0.0002&0.020&0.0006\\
\hline
\end{tabular}
\end{minipage}
\end{table}

Comparing our simulations with the population of known millisecond pulsars,
we find 14 known field millisecond pulsars 
with a known flux density at 1400 MHz in the region of our blind survey, and the faintest one has a
flux density of 0.05 mJy. Down to that value, there are about 33 synthetic
millisecond pulsars, so that about half the total population predicted by our simulations
in that region have already been detected.

Incidentally, the sensitivity of our blind survey is such that, among these 14 known
millisecond pulsars, only \object{PSR B1937+21} and \object{PSR J2007+2722} (which has a period of 24 ms)
can be detected with our setup.

To put the results of our blind survey in perspective, we mention the results of another
blind survey for millisecond pulsars conducted at Nan\c cay at about
the time that we started our own survey.
A survey of 45000 pointings in the region $180^{\circ} \geq l \geq -15^{\circ}$
and $|b| \leq 3^{\circ}$ was conducted in 1997-98 (Foster et al. \cite{foster}),
using a different backend and method of analysis. The frequency, bandpass,
integration time, and nominal sensitivity
of the survey were 1360 MHz, 144 MHz, 120~s, and 0.6 mJy, respectively, and the sampling resolution 
was 1.5 MHz in frequency and 60 $\mu s$ in time.
It produced only two new pulsars, both with periods longer than 100 ms
(Ray et al. \cite{ray}). The reanalysis of 15\% of the stored data with a different
method produced 341 candidate millisecond pulsars, and observing 
200 of them again did not reveal any new pulsar (Desvignes \cite{desvignes}).
With our simulations, we find that there are 30 synthetic millisecond pulsars in
the region of the survey above the flux limit, but only 14 of them are detectable
with our processing method in the best-case scenario (i.e. if the telescope points
exactly at the pulsar in each case).
This confirms what was probably widely believed, that it
is notoriously difficult to detect millisecond pulsars.
As one referee of our telescope time requests stated, it is a high-risk,
high-reward project.

%______________________________________________________________
\section{Conclusions}
\label{conclusions}

The present project was started with no assumptions on the total population of
millisecond pulsars (except that it must be large) and with the specific aim of detecting a population of
fast-spinning (sub)millisecond pulsars.  
The results of the small surveys targeted
at unidentified radiosources suggest that, despite \object{PSR B1937+21}, bright fast-spinning pulsars are rare.
The null result of the blind survey agrees with our present understanding of the field
population of millisecond pulsars, the order of magnitude of the total population, its
absence of strong clustering, and the distribution of periods. 
Our conclusion is that the present models of the Galactic population of millisecond pulsars are not unduly
pessimistic and that the population of fast-spinning millisecond pulsars cannot be unexpectedly large.  

This project was also fruitful in a broader context:
over 24 students worked as interns on the project over the years, and most of them 
subsequently made use of the acquired skills in their career. 

Finally, we hope that
future prospective searches for millisecond pulsars will successfully make use of our experience and
of the numbers given here.

\begin{table*}
\centering
\caption{List of observed faint polarized point-like radiosources}
     \label{tablefs}
  \begin{tabular}{lrrlr|lrrlr}
  \hline
Name               &S1400&\% pol&date&N&Name               &S1400&\% pol&date&N\\
                   &mJy&      &yyyymmdd& &                   &mJy&      &yyyymmdd& \\
  \hline
NVSS J002145+744454& 9.7 & 26.0&20080905&3&NVSS J061712+260152& 7.5 & 27.5&20081103&4\\
NVSS J003706+515737& 7.3 & 40.7&20080905&3&NVSS J070822-151518&10.0 & 26.1&20081204&3\\
NVSS J004419+514533& 2.9 & 84.5&20080905&3&NVSS J071545+051635& 8.5 & 24.9&20081012&4\\
NVSS J012635+704215& 4.4 & 76.8&20080905&4&NVSS J071933-365933& 7.6 & 26.4&20081106&3\\
NVSS J012705+703605& 3.5 & 75.1&20080909&3&NVSS J072438-271618& 5.0 & 43.0&20081106&4\\
NVSS J012818+704655& 5.3 & 64.0&20080909&3&NVSS J072501+004534& 7.5 & 41.6&20081204&4\\
NVSS J012932+701545& 4.2 & 50.5&20080905&4&NVSS J073949+004434& 7.1 & 28.6&20081106&4\\
NVSS J013201+704729& 3.3 &113.3&20080909&4&                   &     &     &20081204&4\\
NVSS J030809+512443& 7.4 & 32.7&20081003&3&NVSS J082008-363214& 6.3 & 34.3&20080703&3\\
NVSS J034707+402127& 9.4 & 28.2&20081003&4&NVSS J082838-380231& 3.4 & 68.2&20080803&4\\
NVSS J035429+643653& 9.5 & 22.5&20081001&3&NVSS J083035-312745& 6.5 & 32.8&20080703&3\\
NVSS J042204+363943& 9.3 & 30.6&20081001&3&NVSS J090355-384439& 3.0 & 67.3&20080703&4\\
NVSS J043739+301858& 5.9 & 66.3&20081001&4&NVSS J155955-382320& 6.0 & 35.3&20080911&3\\
NVSS J050224+250015& 5.1 & 43.3&20081001&4&NVSS J160116-381526& 5.4 & 68.1&20080805&3\\
NVSS J052352+164533& 5.2 & 38.7&20081210&3&NVSS J162707-364109& 3.9 & 55.1&20080805&3\\
NVSS J052906+215305& 6.7 & 59.7&20081212&3&NVSS J162943-340416& 8.4 & 27.3&20080911&3\\
NVSS J053220+163421& 8.8 & 22.8&20081103&3&NVSS J163550-375558& 4.7 & 54.9&20080706&3\\
NVSS J053959+193737& 5.4 & 37.6&20081113&3&NVSS J164751-341010& 4.2 & 48.3&20080805&4\\
NVSS J053959+232629& 2.5 &112.4&20081011&3&NVSS J165146-353221& 5.1 & 39.8&20080706&4\\
NVSS J054134+302901& 9.5 & 22.4&20081210&4&NVSS J165753-305936& 5.1 & 52.2&20080911&4\\
NVSS J054331+500635& 4.1 & 49.5&20081212&4&NVSS J194518+101520& 9.9 & 20.2&20080809&3\\
NVSS J054456+491607& 5.0 & 47.2&20081103&4&NVSS J195016+404408& 4.9 & 44.9&20080807&3\\
NVSS J054532+215524& 8.7 & 24.7&20081011&3&NVSS J200102+420928& 4.6 &100.7&20080807&3\\
NVSS J054650+495147& 6.3 & 44.3&20081113&4&NVSS J200109+112314& 8.4 & 26.8&20080809&3\\
NVSS J054747+392734& 9.3 & 23.9&20081210&4&NVSS J202139+233729& 4.1 & 54.1&20080809&4\\
NVSS J055120+032729& 8.2 & 26.3&20081212&4&NVSS J203453+214620& 4.6 & 88.7&20080807&4\\
NVSS J055600+185916& 9.1 & 27.5&20081103&4&NVSS J204040+304747& 3.5 & 62.9&20080711&3\\
NVSS J060348+215030& 3.6 & 57.2&20081011&4&NVSS J210904+382547& 7.8 & 26.0&20080711&3\\
NVSS J060348+343522&10.0 & 22.0&20081113&4&NVSS J211354+611530& 3.7 & 61.4&20080711&4\\
NVSS J060411+311018& 9.7 & 21.8&20081212&4&NVSS J215324+395318& 8.6 & 24.0&20080711&4\\
NVSS J060820+171726& 8.0 & 26.0&20081210&4&NVSS J223844+451415&10.0 & 20.4&20080708&3\\
NVSS J061011+150106& 7.4 & 28.8&20081011&4&NVSS J230058+713519& 9.8 & 26.4&20080708&3\\
NVSS J061040+332755& 8.6 & 24.1&20081113&4&NVSS J234546+544241& 8.2 & 24.9&20080708&4\\     
\hline
\end{tabular}
\end{table*}

\begin{acknowledgements}
We first wish to acknowledge the help of numerous students who
contributed to this project, mainly on its technical aspects, over the
years: 
Vincent Galeote (1993), J\'er\^ome Moueza (1994-95), Jean-Christophe No\"el (1994-95), 
Yves Rutschl\'e (1995), Eric Soudanas (1995), David Descazaux (1995),
Yannick Magnaud (1995), Olivier Danet (1996-97),
B\'en\'edicte Rougeaux (1995-99), G\'erald Giraud (1996), 
Paul Thierry (1998), Sebastian Karl (1999), Lionel Aub\'epart (2000),
Benoit de Vil\`elle (2001), Mael Guennou (2002), Miguel Charcos (2002), S\'ebastien Rosete (2003),
Franck-Sylvain Petit (2004), Julien Badroudine (2005), Fabien Bricout (2005), 
Angel Boukhari (2005), Alexandre Durand (2006), Amin Shahsavar (2006), and Guillaume Carri\'e (2008). 
This project was supported until 1998 by Centre National d'Etudes Spatiales.
We thank the staff of the Nan\c cay radiotelescope for assistance in
setting up and testing our dedicated backends, and Isma\"el Cognard for help with interfacing
our backend with the telescope and for letting us use the cluster of PCs for data analysis. 
We are grateful to Jean-Michel Martin for
giving us some director's time on the radiotelescope for examining the feasability
of our search for millisecond pulsars in globular clusters, to Pierre Colom for information
on the geometry of the telescope beam, to the anonymous referee for useful comments that
helped improve the paper, and to the language editor for improving the style of the paper.
 
\end{acknowledgements}


\begin{thebibliography}{}
\bibitem[2007]{aulbert} Aulbert,  C. : 2007, MPE Report 291, p.216 (arXiv:astro-ph/0701097)
\bibitem[2010]{bagchi} Bagchi, M., Lorimer, D.R. : 2010, arXiv:1012.4705
\bibitem[1995]{bailyn} Bailyn, C.D., 1995, ARAA, 33, 133 
\bibitem[1997]{cordes97} Cordes, J.M., Chernoff, D.F. : 1997, ApJ 482, 971
\bibitem[2002]{cordes02} Cordes, J.M., Lazio, T.J.W.~: 2002, astro-ph/0207156 
\bibitem[2000]{crawford} Crawford, F., Kaspi, V., Bell, J. : 2000, Pulsar Astronomy, ASP Conf. Ser. 202, 31
\bibitem[2000]{damico} D'Amico N. : 2000, Pulsar Astronomy, ASP Conference Series, Vol. 202, p. 27
\bibitem[2000]{debreuck} De Breuck, C., van  Breugel, W., Rottgering, H.J.A., Miley G. : 2000, A\&AS 143, 303 
\bibitem[2009]{desvignes} Desvignes, G., : 2009, Th\`ese de doctorat de l'Universit\'e d'Orl\'eans
\bibitem[2001]{edwards} Edwards, R.T., van Straten, W., Bailes, M.~: 2001, ApJ 560, 365
\bibitem[2006]{faucher06} Faucher-Gigu\`ere, C.-A., Kaspi, V.M. : 2006, ApJ 643, 332
\bibitem[2010]{faucher10} Faucher-Gigu\`ere, C.-A., Loeb, A. : 2010, JCAP, 1, 5
\bibitem[1997]{foster} Foster, R.S., Ray, P.S., Cadwell, B.J., Backer, D.C., Lestrade, J.-F., Cognard, I., Martin, J.-M., Maitia, V., 1997, BAAS, 29, 1392
\bibitem[1990]{fouque} Fouqu\'e, P., Durand, N., Bottinelli, L., Gouguenheim, L., Paturel, G. : 1990, A\&AS, 86, 473 
\bibitem[2005]{freire} Freire, P.C.C., Hessels, J.W.T., Nice, D.J., Ransom, S.M. : 2005, ApJ 621, 959
\bibitem[2005]{guinot} Guinot, B., Arias, E.F., : 2005, Metrologia, 42, S20 
\bibitem[2003]{guseinov} Guseinov, O.H., Yazgan, E., Tagieva, S.O., \"Ozkan, S. : 2003, RMxAA, 39, 267
\bibitem[1995]{haensel} Haensel, P., Salgado, M., Bonazzola, S.~: 1995, A\&A 296, 745
\bibitem[2004]{han} Han, J.L., Manchester, R.N., Lyne, A.G., Qiao, G.J. : 2004, IAU Symp. 218, 135
\bibitem[2007]{hessels07} Hessels, J.W.T., Ransom, S.M., Stairs, I.H., Kaspi, V.M., Freire, P.C.C. : 2007, ApJ, 670, 363
\bibitem[2007]{hessels08} Hessels, J.W.T., Ransom, S.M., Kaspi, V.M., et al. : 2008, AIPC, 983, 613
\bibitem[1992]{hut} Hut, P., McMillan, S., Goodman, J., et al.~: 1992, PASP 104, 981
\bibitem[2010]{janssen} Janssen, G. H., Stappers, B. W., Bassa, C. G., Cognard, I., Kramer, M., Theureau, G. : 2010, A\&A, 514, 74 
\bibitem[2007]{lattimer} Lattimer, J.M., Prakash, M. : 2007, Phys. Rep. 442, 109 
\bibitem[1995]{lorimer95} Lorimer, D.R., Nicastro, L., Lyne, A.G., et al. : 1995, ApJ, 439, 933 
\bibitem[1996]{lorimer96} Lorimer, D.R., Lyne, A.G., Bailes, M., et al. : 1996, MNRAS, 283, 1383
\bibitem[2006]{lorimer06} Lorimer, D.R., Faulkner, A.J., Lyne, A.G., et al. : 2006, MNRAS, 372, 777 
\bibitem[2007]{lorimer07} Lorimer, D.R., McLaughlin, M.A., Champion, D.J., Stairs, I.H. : 2007, MNRAS, 379, 282 
\bibitem[2008]{lorimer08} Lorimer, D.R. : 2008, Living Reviews in Relativity, 11, 8
\bibitem[1985]{lyne85} Lyne, A.G., Manchester, R.N., Taylor, J.H. : 1985, MNRAS, 213, 613
\bibitem[1987]{lyne87} Lyne, A.G., Brinklow, A., Middlewitch, J., Kulkarni, S.R., Backer, D.C., Clifton, T.R.: 1987, Nature 328, 399
\bibitem[1998]{lyne98} Lyne, A.G., Manchester, R.N., Lorimer, D.R., et al. : 1998, MNRAS, 295, 743 
\bibitem[2005]{M05} Manchester, R.N., Hobbs, G.B., Teoh, A., Hobbs, M. : 2005, AJ 129, 1993
\bibitem[2011]{M11} Manchester, R.N. : 2011, arXiv:1101.5202 
\bibitem[1995]{navarro} Navarro, J., De Bruyn, A.G., Frail, D.A., Kulkarni, S.R., Lyne, A.G. : 1995, ApJ 455, L55
\bibitem[1996]{petit} Petit, G., Tavella, P.~: 1996, A\&A 308, 290
\bibitem[2003]{possenti} Possenti, A., D'Amico, N., Manchester, R.N., Camilo, F. : 2003, ApJ 599, 475
\bibitem[1999]{ray} Ray, P.S., Cadwell, B.J., Lazio, T.J.W., Foster, R.S., Backer, D.C., Cognard, I., Lestrade, J.-F., 1999, BAAS, 31, 903
\bibitem[2005]{ransom} Ransom, S.M., Hessels, J.W.T, Stairs, I.H., et al. : 2005, Science 307, 892
\bibitem[1999]{BR} Rougeaux B.~: 1999, Th\`ese de doctorat de l'Universit\'e Paul Sabatier
\bibitem[2000]{BR2000} Rougeaux, B., Petit, G., Fayard, T., Davoust E.~: 2000, Experimental Astronomy 10, 473
\bibitem[2010]{shi} Shi, H., Liang, H., Han, J. L., Hunstead, R. W. : 2010, MNRAS, 409, 821 
\bibitem[1996]{stappers} Stappers, B.W., Bailes, M., Lyne, A.G., Manchester, R.N., D'Amico, N., et al. 1996, ApJ, 465, L119
\bibitem[2007]{story} Story, S., Gonthier, P.L., Harding, A.K. : 2007, ApJ 671, 713
\bibitem[1991]{taylor} Taylor, J.H. : 1991, Proc. IEEE, 79, 1054 
\bibitem[2009]{vanhaasteren} van Haasteren, R., Levin, Y., McDonald, P., Lu, T. : 2009, MNRAS, 395, 1005 
\bibitem[2009]{verbiest} Verbiest, J.P.W., Bailes, M., Coles, W.A., et al. : 2009, MNRAS 400, 951
\bibitem[2011]{yan} Yan, W.M., Manchester, R.N., van Straten, W., et al. : 2011, MNRAS, in press (arXiv:1102.2274)
\bibitem[2004]{yusifov} Yusifov, I. : 2004, in {\it The Magnetized Interstellar Medium}, ed. B. Uyaniker et al., p.165
\bibitem[2004]{yusifov04} Yusifov, I., K\"u\c c\"uk, I. : 2004, A\&A, 422, 545
\end{thebibliography}
\end{document}